\documentclass[journal]{IEEEtran}
\ifCLASSINFOpdf
\else
\fi

\usepackage{cite}
\usepackage{amsmath,amssymb,amsfonts}
\usepackage{algorithmic}
\usepackage{graphicx}
\usepackage{textcomp}
\usepackage{xcolor}
\usepackage{rotating}
\usepackage{epsfig,graphicx,lscape}
\usepackage{amssymb}
\usepackage{amsmath} 
\usepackage{amsbsy}
\usepackage{float}
\usepackage{multirow}
\usepackage{array}
\usepackage{marginnote}

\newcommand{\T}{\mathrm{T}}

\newcommand{\xvect}{\mathbf{x}}
\newcommand{\yvect}{\mathbf{y}}

\newcommand{\rvect}{\mathbf{r}}
\newcommand{\wvect}{\mathbf{w}}
\newcommand{\alphavect}{\boldsymbol{\alpha}}
\newcommand{\betavect}{\boldsymbol{\beta}}

\newcolumntype{L}[1]{>{\raggedright\let\newline\\\arraybackslash\hspace{0pt}}m{#1}}
\newcolumntype{C}[1]{>{\centering\let\newline\\\arraybackslash\hspace{0pt}}m{#1}}
\newcolumntype{R}[1]{>{\raggedleft\let\newline\\\arraybackslash\hspace{0pt}}m{#1}}

\def\BibTeX{{\rm B\kern-.05em{\sc i\kern-.025em b}\kern-.08em
    T\kern-.1667em\lower.7ex\hbox{E}\kern-.125emX}}

\begin{document}
%\title{Towards Understanding Emotional Experience in a Componential Framework}
\title{A Multi-Componential Approach to Emotion Recognition and the Effect of Personality }

\author{Gelareh~Mohammadi,~\IEEEmembership{Member,~IEEE,}
        Patrik~Vuilleumier% <-this % stops a space
\thanks{G. Mohammadi is with the School of Computer Science and Engineering, UNSW Sydney, Australia e-mail: g.mohammadi@unsw.edu.au.}% <-this % stops a space
\thanks{P. Vuilleumier is with the Swiss Center for Affective Sciences, University of Geneva, Switzerland e-mail: patrik.vuilleumier@unige.ch }% <-this % stops a space
}


\markboth{IEEE Transactions on Affective Computing}%
{Shell \MakeLowercase{\textit{et al.}}: Bare Demo of IEEEtran.cls for IEEE Journals}

\IEEEoverridecommandlockouts
\IEEEpubid{1949-3045~\copyright~2020 IEEE. Personal use is permitted, but republication/redistribution requires IEEE permission.}


\maketitle

\begin{abstract}
Emotions are an inseparable part of human nature affecting our behavior in response to the outside world. Although most empirical studies have been dominated by two theoretical models including discrete categories of emotion and dichotomous dimensions, results from neuroscience approaches suggest a multi-processes mechanism underpinning emotional experience with a large overlap across different emotions. While these findings are consistent with the influential theories of emotion in psychology that emphasize a role for multiple component processes to generate emotion episodes, few studies have systematically investigated the relationship between discrete emotions and a full componential view. This paper applies a componential framework with a data-driven approach to characterize emotional experiences evoked during movie watching.  The results suggest that differences between various emotions can be captured by a few (at least 6) latent dimensions, each defined by features associated with component processes, including appraisal, expression, physiology, motivation, and feeling. In addition, the link between discrete emotions and component model is explored and results show that a componential model with a limited number of descriptors is still able to predict the level of experienced discrete emotion(s) to a satisfactory level. Finally, as appraisals may vary according to individual dispositions and biases, we also study the relationship between personality traits and emotions in our computational framework and show that the role of personality on discrete emotion differences can be better justified using the component model.
\end{abstract}


% \fi
\IEEEpeerreviewmaketitle

\section{Introduction}
\IEEEPARstart{E}{motions} are inseparable part of human nature, at the centre of every social processes, which not only affect the feeling state but also shape one's perception~\cite{Phelps2006}, cognition~\cite{Dolan2006}, action\cite{damasio2006descartes} and memory~\cite{Phelps2004, Tambini2017}. Therefore, the ability to recognise emotion in others and respond appropriately is vital to maintain any relationships\cite{Ekman2004, Levine2007}. Despite the great efforts in conceptualising emotion experience, various theories have remained debated~\cite{Sander2005}. However, there is a general consensus that emotions are multi-componential phenomena consisting of appraisal of an event followed by motivation to take action(s), motor expression in the face and body, as well as changes in physiology and subjective feeling\cite{Mortillaro2015}. Nevertheless, most of the previous works on emotion recognition or neural circuitry of emotion have mainly focused on changes in the feeling component either in the form of discrete emotions or dimensional model\cite{Wu2014, Koolagudi2012, Konar2014}. Discrete model of emotion postulates a small set of basic emotions, shared across cultures, that each represents a distinct feeling with a unique facial expression\cite{Mortillaro2015}. The most popular set of basic emotions was introduced by Ekman and his colleague which has six basic emotions of anger, disgust, fear, happiness, sadness and surprise~\cite{Ekman2011,Ekman1992, Weninger2015}. In contrast, dimensional model of emotion describes any feeling state according to one or more dimensions like valence and arousal~\cite{Russell2003}. Although such models of emotion are very useful in many areas, they neglect the complexity of emotions altogether, and reduce an emotional experience to either a fixed label or a point in valence-arousal space. Feeling component, as an integrated awareness of the changes in other components, holds an exceptional position in componential model of emotion, however, it doesn't represent the involved processes which led to that awareness. Therefore, to better understand the underpinning mechanism of different emotional states, it is important to consider the full componential view. Furthermore, by considering the effect of different appraisals in the elicitation of emotion, this frameworks allows for more precise assessment of individual differences and personality-related biases that may determine different emotions to similar events in different people ~\cite{Scherer2009}.

\IEEEpubidadjcol 

In this paper, we first present a dataset, collected by inducing a wide range of emotions in an effort to span the componential space, well. Then we demonstrate what we learn about the underlying dimensions of emotional experience using unsupervised learning methods. And finally, we present computational analyses used to study the link between discrete emotions and componential model of emotion and analyse the importance of different descriptors in component model. We have also looked into the potential effect of personality traits on experienced emotion. The main contributions of this work are as follows: first, this study is among the first works that investigate the mapping between discrete emotions and the full componential model using a data-driven approach; second, most previous studies have only focused on empirical assessments of discrete emotions in terms of semantic profiles rather than reporting on actual emotional experience which is the focus of this work; third, we examine the importance of different components as well as different descriptors in predicting discrete emotions; and finally, this work goes beyond just labelling each profile with one discrete emotion and instead predicts the degree to which each discrete emotion is felt.

The current manuscript is an extension to the previous work presented in the \emph{International Conference on Affective Computing and Intelligent Interaction}~\cite{Mohammadi2019} and introduces more comprehensive analyses and evaluations, including a relationship analysis between personality and affects to better understand how personality modulates the experienced emotion.

\IEEEpubidadjcol

\section{Background}

\subsection{Component Process Model}
Component process model (CPM) is an alternative view to explain emotions which borrows elements of dimensional models, as emergent results of underlying dimensions, and elements of discrete model, that postulates different subjective qualities, and enriches those by adding a layer of cognitive mechanism at the basis of emotional experience. This model assumes a series of cognitive evaluation on different levels of processing which introduces more flexibility and can account for individual and contextual differences in emotional responses to the same stimuli. Furthermore, such model can accomodate the non-prototypical and mixed/fuzzy emotion categories~\cite{Mortillaro2012}, as well.

\begin{figure}[!t]
\begin{center}
\includegraphics[width=9cm]{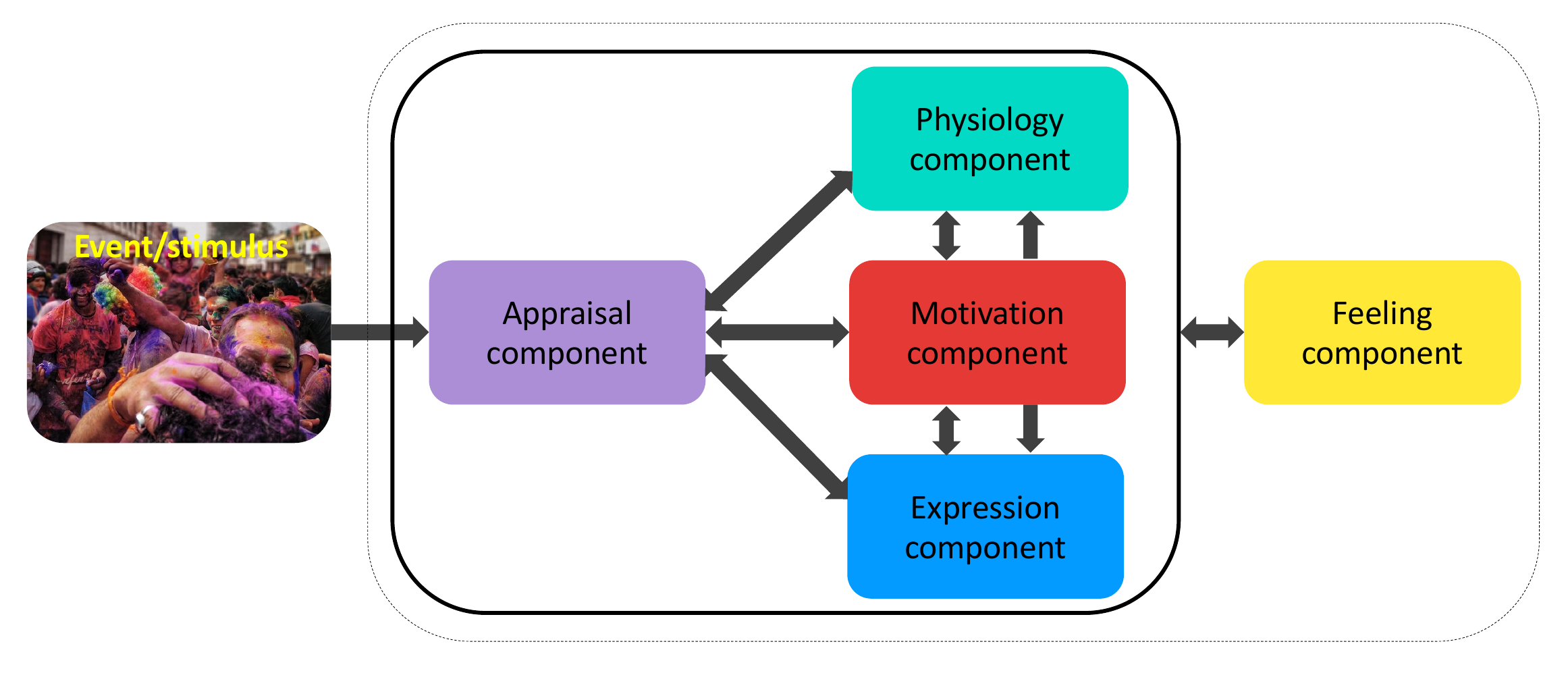}
\end{center}
\caption{Component Model of Emotion with five components, as suggested by Scherer; It starts by \emph{Appraisal} of the event which leads to changes in \emph{Physiology}, \emph{Motivation} and \emph{Expression} components, and changes in all those four components bring us to a \emph{Feeling} state.}\label{ComponentModel}
\end{figure}

According to component process model of emotion, the subsystem changes in an organism are driven by parallel appraisals which reflect the subjective assessment of an event and activates changes in other subsystems. In this model, every emotional experience arises from coordinated changes in five components which starts with 1) the appraisal component that implies evaluating the event/situation with respect to personal significance, implications for goals, coping potential, novelty, and compatibility with norms; the evaluations are drawn from memory and goal representations and examine the relevance of the event and its consequence for the organism's well-being. The changes in this component trigger changes in the other four main components including: 2) a motivation component that defines changes in action tendencies (e.g. fight, flight or freeze) and prepares for appropriate response(s); 3) a physiological component that encompasses changes in peripheral autonomic activity (e.g. changes in heart rate or respiratory rate) and occurs based on the appraisal results and associated motivational changes; 4) an expression component that involves changes in expressive motor behavior such as facial expression, body gestures/postures and also gets activated similar to physiology component as a result of the appraisal and motivation components; and finally 5) a feeling component that reflects the conscious experience associated with changes in all other components, usually described by people with some categorical labels (like anger, happiness, sadness and so on)~\cite{Scherer2009}. (see Figure~\ref{ComponentModel})

CPM defines a series of cognitive assessments which can potentially predict which emotion would be experienced based on the appraisals. The appraisals can be seen as the causal mechanism for modifying action tendencies and then these two components together are the basis of changes in physiology and expression components. The relationship between different discrete emotions and the underlying components are studied in two ways of theory-driven and data-driven modelling; in theory-driven models, the componential emotion theorists propose a particular profile for each discrete emotion (a top-down model) and they specify which combination of appraisal criteria produces a certain discrete emotion, usually, from a limited set of emotions~\cite{Frijda1986, Lazarus2001}. One important assumption in theory-driven approaches is the interaction between appraisal criteria which typically imposes a hierarchical structure in which different criteria interact in a decision-tree sort of architecture. A main criteria appears at the top of the tree and separates major categories of emotions followed by interactions at lower levels that differentiate more specific categories of emotions. In most of such theories, the \emph{relevance} criteria appears at the top level and is an indicator of whether any emotion has been elicited or not. The most dominant criteria after relevance is goal compatibility which examines the event in terms of its impact on one's goals and desire and therefore is assumed to differentiate between positive and negative emotions. 

In contrast, data-driven models use data to define the link between discrete emotion and their representation in componential space (a bottom-up model). Here, the idea is to collect data on emotional experience and let the computational models to detect the relation between discrete emotions and their componential representations. The main advantage of such approaches is that they posit fewer assumptions on the nature of the relationship and let the data to determine that.  Most of the data-driven approaches have only focused on the relationship between discrete emotions and appraisal component and only very few works have looked into the full componential process model. Some of these studies have focused on discriminative power of appraisal component for discrete emotion categories and their reported accuracies varies between 27\% to 80\% depending on the number of emotion categories they have taken into account~\cite{Fontaine2013,Frijda1989,Reisenzein1994,Ruth2002,Meuleman2013}. Other studies investigated the link between specific emotions and specific appraisals by either manipulating the appraisal\cite{McGraw1987, Russell1986, Stipek1989} or observed data\cite{Kuppens2003, Scherer1997, Siemer2007}. There is only one recent study that has taken similar approach with a different experimental setup, using virtual reality for emotion elicitation\cite{Meuleman2018}. Participants played 7 games and evaluated their experience in terms of the intensity for 20 qualitative discrete emotion categories of the Geneva Emotion Wheel (GEW) and 39 descriptors from CoreGRID questionnaire (see section material and assessment). However, further analyses revealed that only fear and joy are involving a broad pattern of component responses and can be used in the subsequent modelling. Therefore, only two categories of fear and joy are modelled as a function of a set of component model descriptors. They applied a multilevel models using forward stepwise modelling. For fear, the best model with 9 descriptors has achieved a marginal $R^2$ of 0.62 and conditional $R^2$ of 0.69 and for joy the best model is achieved with 3 predictors, resulting a marginal $R^2$ of 0.26 and conditional $R^2$ of 0.66. The main limitation of that study was the lack of variety in the elicited emotions and appraisal criteria.

\subsection{Personality}
Personality is defined as a latent construct accounting for \emph{``individuals characteristic patterns of thought, emotion, and behavior across various situations, together with the psychological mechanisms - hidden or not - driving those patterns"}~\cite{Funder2001}. Different models of personality has been proposed to capture these latent traits~\cite{Saucier1996}, however the most widely used representation is the \emph{Big Five} (BF) model which captures the most phenotypic individual differences through five major factors~\cite{Matthews2003} corresponding to:
\begin{itemize}
\item\emph{Extraversion}: Active, Assertive, Energetic, etc.%etc,%, Talkative
\item\emph{Agreeableness}: Appreciative, Kind, Generous, etc. %,  Forgiving, Sympathetic, 
\item\emph{Conscientiousness}: Efficient, Organized, Planful, etc.%, Reliable%, , Thorough
\item\emph{Neuroticism}: Anxious, Self-pitying, Tense, Unstable, etc.%Touchy, etc.%, , Worrying
\item\emph{Openness}: Artistic, Curious, Imaginative, etc.%Insightful, etc.%, Original, 
\end{itemize}
In the BF model, personality is represented with five scores, each corresponding to one of the traits. These scores are measured through some standard questionnaires and in this work we have used \emph{Big Five Inventory 10} (BFI-10), which is a short personality questionnaire with 10 items selected from the full BFI questionnaire (44 items) based on their higher correlation with the  personality trait scores~\cite{Rammstedt2007}. 

Although the main focus of this study is on the organisation of discrete emotion in componential space, we were also interested to look at the potential effect of personality on experienced emotions to see whether personality modifies the experienced emotion. Individual differences and biases may influence the engagement of different appraisal processes in different people and thus contribute to generate different affective responses in different people. 

\subsection{Personality and Emotion}
People are different in the way they perceive and react to emotional information, but such differences are not solely captured by their personality characteristics. Nevertheless, such dispositions have the potential to affect our emotional behavior; and therefore, it is important to consider personality differences to better understand why people differ in their emotional reactions. The association between personality and emotion has been largely studied in social psychology and several studies have shown that personality traits affect our emotional experience~\cite{Steel2008}. 
Such associations are important since they can reveal the link between personality traits and psychopathology and potentially explain the mechanism by which certain characteristics of personality can lead to higher rates of psychopathology. For example, experiencing higher levels of negative affect and being more reactive to internal/external stressors can increase the likelihood of certain psychological disorders~\cite{Komulainen2014}, and chronic cognitive biases can favor maladaptive responses to stressful situation and thus contribute to the emergence of mood and anxiety disorders~\cite{Zautra2005}. 

Most of such studies have mainly looked at the relationship between personality and dimensional model of emotion~\cite{Winter1997}, but their findings are sometimes in contrast to each other. For example~\cite{Kehoe2012} found that \emph{Neuroticism} is negatively correlated with positive valence, but  \cite{Ng2009} observed a positive correlation between \emph{Neuroticism} and valence in negative stimuli, with no significant correlation in positive stimuli. Similarly, \cite{Kehoe2012} , \cite{Stenberg1992} and \cite{Pascalis2004} all reported positive correlated between \emph{Neuroticism} and arousal, however, \cite{Stenberg1992} and \cite{Pascalis2004} showed that the correlation is stronger for negative-valence stimuli. Most studies have associated \emph{Extraversion} with positive valence~\cite{Zautra2005, Lucas2000, Costa1980, Kuppens2007} . A negative correlation between \emph{Extraversion} and arousal was noted in \cite{Stenberg1992} which was also reported in \cite{Gupta1985}, though \cite{Brumbaugh2013} reported positive correlation between both \emph{Extravert} and \emph{Neuroticism} with arousal. \emph{Agreeableness} and \emph{Conscientiousness} have been linked to higher positive affect \cite{Kuppens2007, DeNeve1998, Steel2008, Tong2006} and although several studies have reported no significant correlation between  \emph{Openness to experience} and positive or negative affect~\cite{Kuppens2007, Miller2006}, a meta-analysis has associated this trait with higher positive affect~\cite{Steel2008}.

The role of personality in appraisal biases has also been considered in multiple studies~\cite{Scherer2009a}. However, datasets for studying personality-appraisal relationships quantitatively are scarce. In one such study~\cite{Tong2006}, \emph{Neuroticism} was shown to be negatively correlated with perceived control and certainty, and positively correlated with Unfairness and moral violation (at $p<0.01$) . \emph{Extraversion}, \emph{Agreeableness} and \emph{Openness} did not show any significant correlation with any of appraisals (at $p<0.01$). \emph{Conscientiousness} showed significant positive correlation with perceived control and negative correlation with unfairness and moral violations (at $p<0.01$). The role of personality on motivation has also been postulated in some studies~\cite{Griner2000}, however we could not find any work that have considered the full componential framework.

Therefore, while multiple studies have looked into such associations, the results are somewhat mixed or insufficient and further analyses are necessary to better understand how personality traits affect our emotional experience. So, one aim of this study is to explore the relationship between personality traits and different emotions in the perspective of appraisal processes that are postulated by componential theories of emotion. According to the results from majority of literature, our initial hypotheses are that \emph{Extraversion}, \emph{Agreeableness}, \emph{Conscientiousness} and \emph{Openness} are positively correlated with positive affect, whereas, \emph{Neuroticism} is positively associated with negative affect and negatively with positive affect. We also assume that the effect of personality on discrete emotions can not be fully explained by the relationship between personality and valence/arousal dimensions and considering the full componential model might be essential in justifying the effect.  

\section{Approach}

\subsection{Material \& Assessment}

The first step to analyse emotions from CPM perspective is to elicit a wide variety of emotions to span the componential space as much as possible. To elicit different emotions we made use of film excerpts. The use of film excerpts for emotion elicitation has been well established in empirical studies of affects due to their desirable characteristics including being dynamic, readily standardized, accessible and ecologically valid \cite{Gross1995}. Several studies have already shown the efficacy of film stimuli in inducing different emotions\cite{Gross1995, Schaefer2010, Samson2016, Gabert-Quillen2015}. Moreover, films are considered as naturalistic stimuli which can induce even complex emotions like nostalgia and empathy\cite{Raz2013} or mixed emotions.

To select a set of emotionally engaging film excerpts a collection of 139 video clips from the well-known literatures on emotion elicitation was selected\cite{Gross1995, Schaefer2010, Gabert-Quillen2015, Soleymani2008} based on availability. The emotion assessment was done in terms of discrete emotions and componential model descriptors. We used a modified version of Differential Emotion Scale to evaluate 14 discrete emotions namely fear, anxiety, anger, shame, warm-hearted, joy, sadness, satisfaction, surprise, love, guilt, disgust, contempt and calm\cite{Izard1993, McHugo1982}. For component model we used a questionnaire with 39 descriptive items which is a subset of CoreGRID instrument with 63 items representing activity in all five major components\cite{Fontaine2013}. The item selection was performed based on the applicability to the emotion elicitation scenario which is a passive emotional response to an event in a video clip, rather than an active involvement in a situation. For example items such as ``it was caused by my own behavior" or ``it was important and relevant for my goals" were removed from the list as the answer to such items would always be "No" in this scenario and therefore they will be non-informative to the analyses  due to absence of any variance. We also made sure that the collected items represent all five major components (appraisal, motivation, expression, physiology and feeling). Please note that removing items will not affect the psychometric properties of the CoreGRID instrument, as the tool is not developed to measure any construct and it is rather used to represent the semantic profile of each emotion across different components. Table 1 summarises the items used in the experiment along with BFI-10 personality questionnaire and the list of discrete emotions that were assessed. 
\begin{table}[!htbp]
\begin{center}
\footnotesize
\begin{tabular}{cll}
 \hline
& Big Five Inventory 10 (BFI-10) &\\
\hline
 1 & This person is reserved &\\
2 & This person is generally trusting &\\
3 & This person tends to be lazy &\\
4 & This person is relaxed, handles stress well &\\
5 & This person has few artistic interests &\\
6 & This person is outgoing, sociable &\\
7 & This person tends to find fault with others &\\
8 & This person does a thorough job &\\
9 & This person gets nervous easily &\\
10 & This person has an active imagination &\\
\hline
\hline
& GRID Questionnaire & Component \\
\hline
 & While watching this movie, did you... & \\
1 & think it was incongruent with your standards/ideas? & Appraisal\\
2 & feel that the event was unpredictable ? & Appraisal\\
3 & feel the event occurred suddenly? & Appraisal\\
4 & think the event was caused by chance? & Appraisal\\
5 & think that the consequence was predictable? & Appraisal\\
6 & feel it was unpleasant for someone else? & Appraisal\\
7 & think it was important for somebody?s goal or need? & Appraisal\\
8 & think it violated laws/social norms? & Appraisal\\
9 & feel in itself was unpleasant for you? & Appraisal\\
10 & want the situation to continue? & Motivation\\
11 & feel the urge to stop what was happening? & Motivation\\
12 & want to undo what was happening? & Motivation\\
13 & lack the motivated to pay attention to the scene? & Motivation\\
14 & want to destroy s.th.? & Motivation\\
15 & want to damage, hit or say s.th. that hurts? & Motivation\\
16 & want to tackle the situation and do s.th.? & Motivation\\
17 & have a feeling of lump in the throat? & Physiology\\
18 & have stomach trouble? & Physiology\\
19 & experience muscles tensing? & Physiology\\
20 & feel warm? & Physiology\\
21 & sweat? & Physiology\\
22 & feel heartbeat getting faster? & Physiology\\
23 & feel breathing getting faster? & Physiology\\
24 & feel breathing slowing down? & Physiology\\                         
25 & produce abrupt body movement? & Expression \\
26 & close your eyes? & Expression\\
27 & press lips together? & Expression\\
28 & have the jaw drop? & Expression\\
29 & show tears? & Expression\\
30 & have eyebrow go up? & Expression\\
31 & smile? & Expression\\
32 & frown? & Expression\\
33 & produce speech disturbances? & Expression\\
34 & feel good? & Feeling\\
35 & feel bad? & Feeling\\
36 & feel calm? & Feeling\\
37 & feel strong? & Feeling\\
38 & feel an intense emotional state? & Feeling\\
39 & experience an emotional state for a long time? & Feeling\\
\hline
\hline
& Discrete Emotions & \\
\hline
& While watching this movie, did you feel... & \\
1 & fearful, scared, afraid? & \\
2 & anxious, tense, nervous? & \\
3 & angry, irritated, mad? & \\
4 & warm, hearted, gleeful, elated? &  \\
5 & joyful, amused, happy?  & \\
6 & sad, downhearted, blue? & \\
7 & satisfied, pleased? & \\
8 & surprised, amazed, astonished? & \\
9 & loving, affectionate, friendly? & \\
10 & guilty, remorseful? & \\
11 & disgusted, turned off, repulsed? & \\
12 & disdainful, scornful, contemptuous? & \\
13 & calm, serene, relaxed? & \\
14 & ashamed, embarrassed? & \\
\hline
& &
\end{tabular}
\caption{The questionnairused in the experiment of this work which included three parts including assessment of personality, GRID features and discrete emotions}
\label{questions}
\end{center}
\end{table}

\subsection{Experimental Setup}

The assessment was done through a web interface using CrowdFlower, a crowdsourcing platform which allows accessing an online workforce to perform a task. The selected workforce was limited to native English speakers from USA or UK. Every task was estimated to take on average 10 min according to a pilot test and the reward was set at 10\textcent (USD) per task (an effective hourly wage of \$6.6 (USD)). The quality control of the assessments was taken care of by means of six test questions about the content of the clip. For example, participants had to indicate whether they have seen an animal in the scene, whether there was any sport activity or whether they heard a gunshot while watching the clip. These questions were designed to ensure that participants had watched the clip thoroughly and if there were more than 4 wrong answers, the data  has been discarded.

\begin{figure}[!t]
\begin{center}
\includegraphics[width=8.6cm]{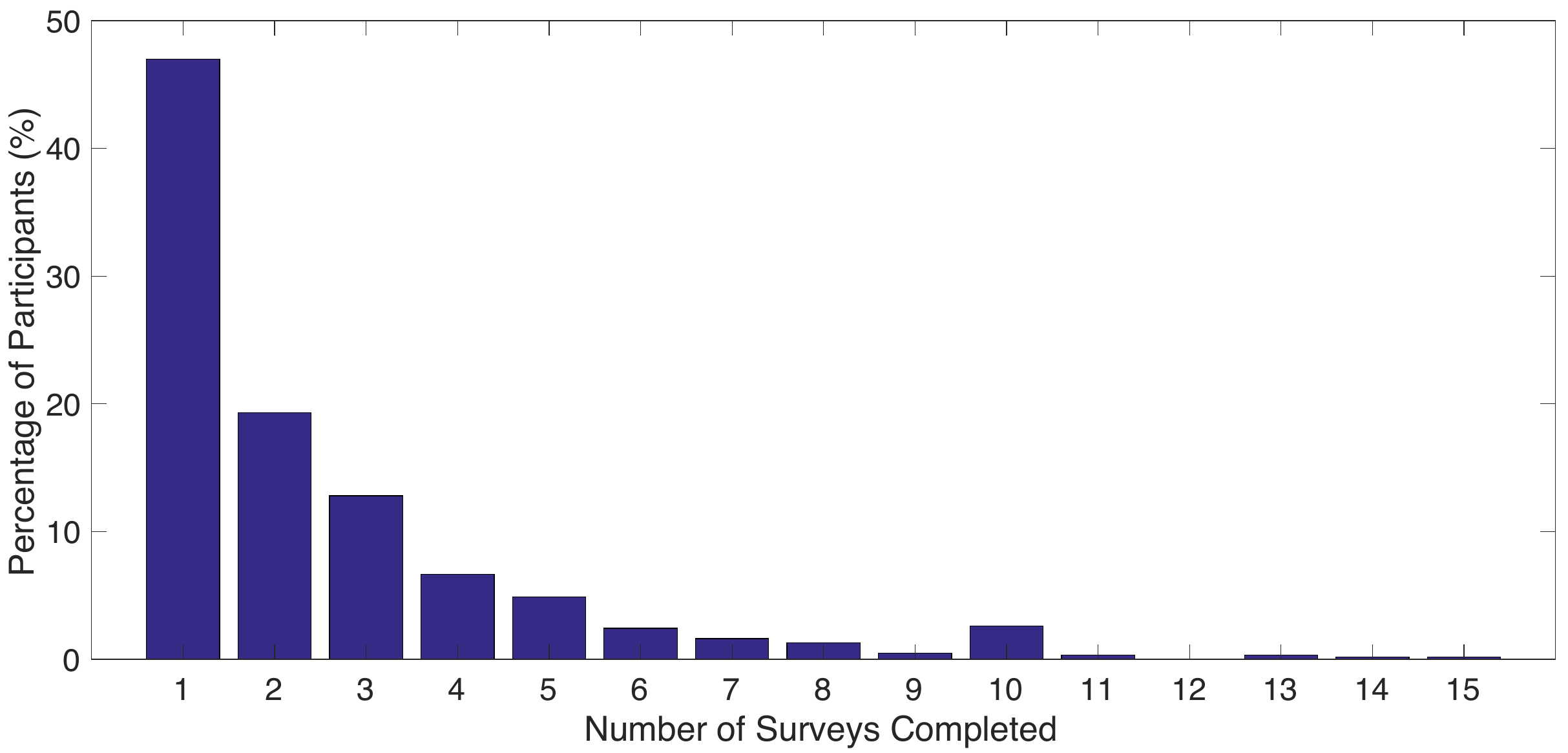}
\end{center}
\caption{Distribution of number of surveys completed per participant. The chart shows the percentage of participants completing the survey a given number of times.}\label{workersHistogram}
\end{figure}

\begin{figure}[!t]
\begin{center}
\includegraphics[width=8.8cm]{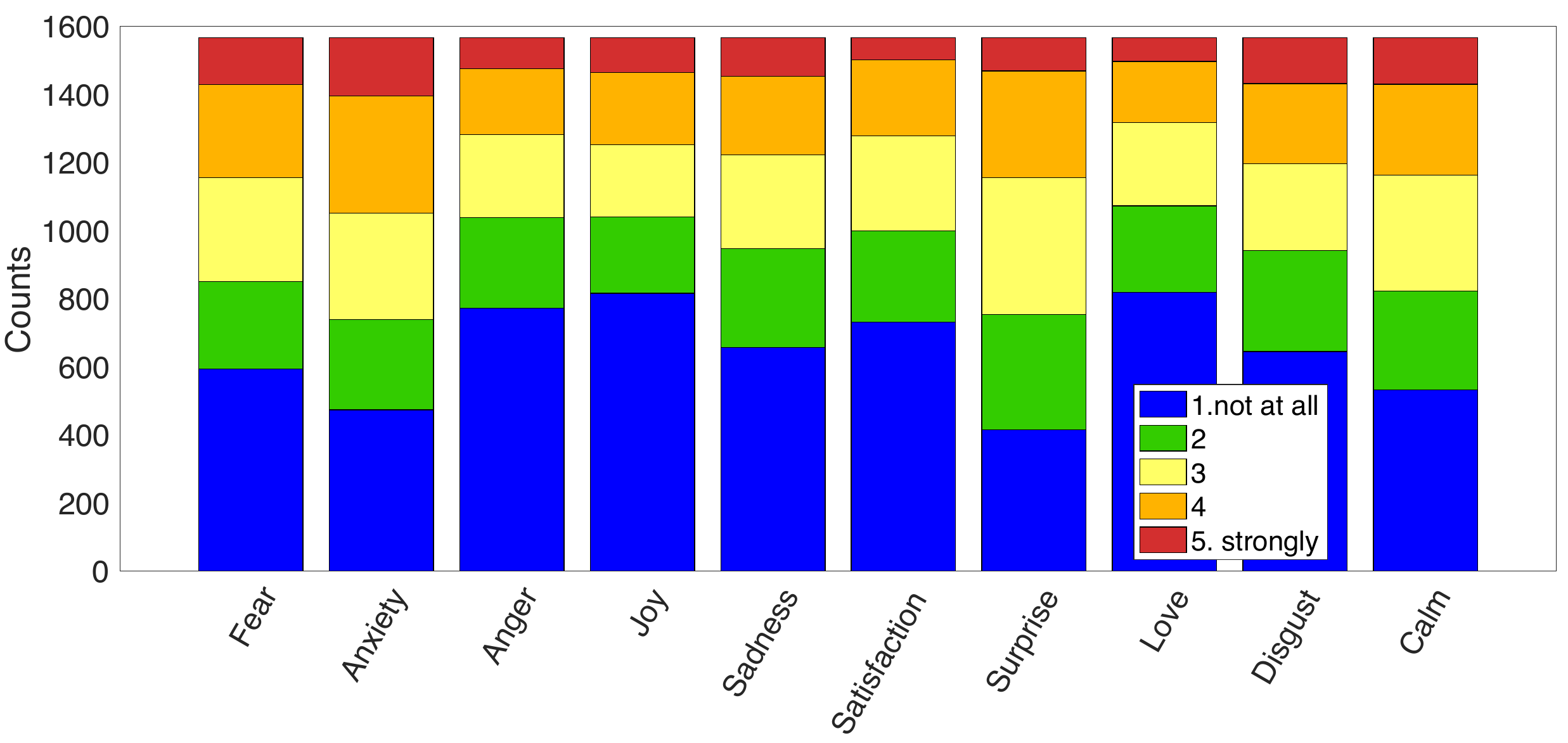}
\end{center}
\caption{Distribution of rankings in each discrete emotion. Each colour shows the proportion of samples with corresponding ranking.}\label{DiscreteEmotionDistribution}
\end{figure}

Participants were required to first self-assess their personality using the Big Five Inventory 10 (BFI-10) questionnaire~\cite{Rammstedt2007}, then watch the video clip and finally completing the GRID and discrete emotion questionnaires by rating how much each question described their feeling or experience on a 5-points likert-scale with 1 associated to ``not at all" and 5 associated to ``strongly''. Participants were instructed to let themselves to freely feel the emotions and express them rather than controlling the feelings and then reflect on what they felt in the assessment rather than ``what they think they should feel". In a pilot study, a set of 5 assessments was collected per each clip and 99 video clips from the original set were selected based on the intensity of induced emotion(s) and the emotion discreteness. In the second round of assessments, 10 more judgments were collected for each clip to provide a higher statistical power for inference of emotionality. No clip with high ratings for shame, warm-hearted, guilt and contempt were found, so these four emotions were excluded from the list of elicited emotions. The total duration of the dataset is about 3.7 hours with an average length of 133.8 seconds for each clip.  Overall 1792 validated survey results from 638 workers (358 males, mean-age = 34, SD = 11 )were collected from which 1567 surveys belonged to the 99 selected clips with at least 15 assessments per video clip, gathered from 617 participants (workers with unique ID). Figure~\ref{workersHistogram} show the distribution of number of surveys completed per worker. To evaluate the power of selected video clips in inducing a wide range of emotions,  Figure~\ref{DiscreteEmotionDistribution} represent the portion of samples within each ranking group across all discrete emotions. Overall, the ratings cover the whole range of the continuum, however it is more skewed towards the lower end particularly for anger, joy, satisfaction and love.

\section{Analysis \& Results}

\subsection{Cluster Analysis}

\begin{figure}[!t]
\begin{center}
\includegraphics[width=9cm]{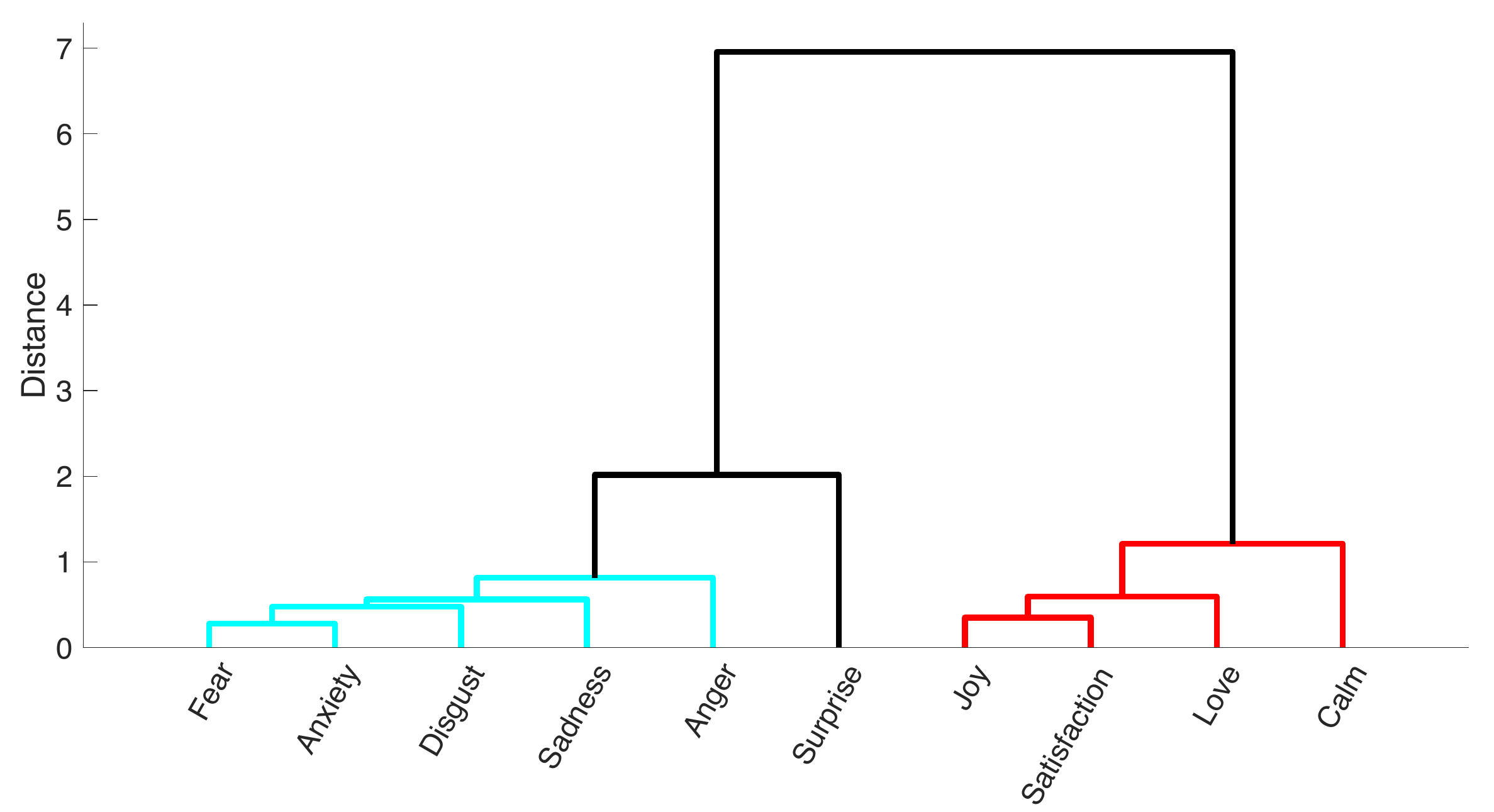}
\end{center}
\caption{Clustering of discrete emotion profile in componential space using hierarchical clustering with squared Euclidean distance. In this method, emotions with more similarity appear closer in the dendrogram and the distance value at each join represents the dissimilarity between the two branches of that split. In this graph, negative emotions appear as a cluster in the left and positive emotion as a cluster in the right side of the dendrogram. }\label{ClusterAnalysis}
\end{figure}

To investigate the similarity between different discrete emotions in componential space, we ran a clustering analysis on the componential profile of discrete emotions. The componential profile of each discrete emotion was estimated using a weighted average of normalised GRID-feature ratings where the weight for each observation is proportional to its rating for that specific discrete emotion. For example, if there is a set of $n$ ratings for GRID features $\{\rvect_{1},\ldots,\rvect_{n}\}$, where each vector is described by $d=39$ ratings ($\rvect_i \in \mathbb{R}^{39} $) and a corresponding set of $n$ ratings for discrete emotions $\{\wvect_{1},\ldots,\wvect_{n}\}$ where each vector is described by $d=10$ ratings  ($\wvect_i \in \mathbb{R}^{10} $) and $\wvect_{ij}$ is the rating for discrete emotion $j$ in sample $i$, the componential profile ($CP$) for discrete emotion $j$ is defined as:
\begin{equation}
	CP_{j} = \dfrac{\sum_{i=1}^{n} \wvect_{ij} \xvect_{i}}{\sum_{i=1}^{n} \wvect_{ij}}
\end{equation}
Each $CP_{j}$ is a vector representing the class centroid corresponding to one discrete emotions $j$. A hierarchical clustering analysis based on Ward's method, which applies an agglomerative hierarchical clustering procedure over squared Euclidian distance, was performed on the class centurions\cite{Murtagh2014}. At the high-level, our results suggest a clear distinction between positive versus negative emotions, while five main clusters are observed at the lower level (Figure~\ref{ClusterAnalysis}). These included clusters for happiness (joy, satisfaction, love), serenity (calm), distress (fear, anxiety, disgust, sadness), anger, and surprise. These findings demonstrate that theoretically meaningful clusters were deriven from our set of features, and thus validates the experimental paradigm and its success in eliciting different emotion categories with expected characteristics.

\begin{figure*}[!htbp]
\begin{center}
\includegraphics[width=17.5cm]{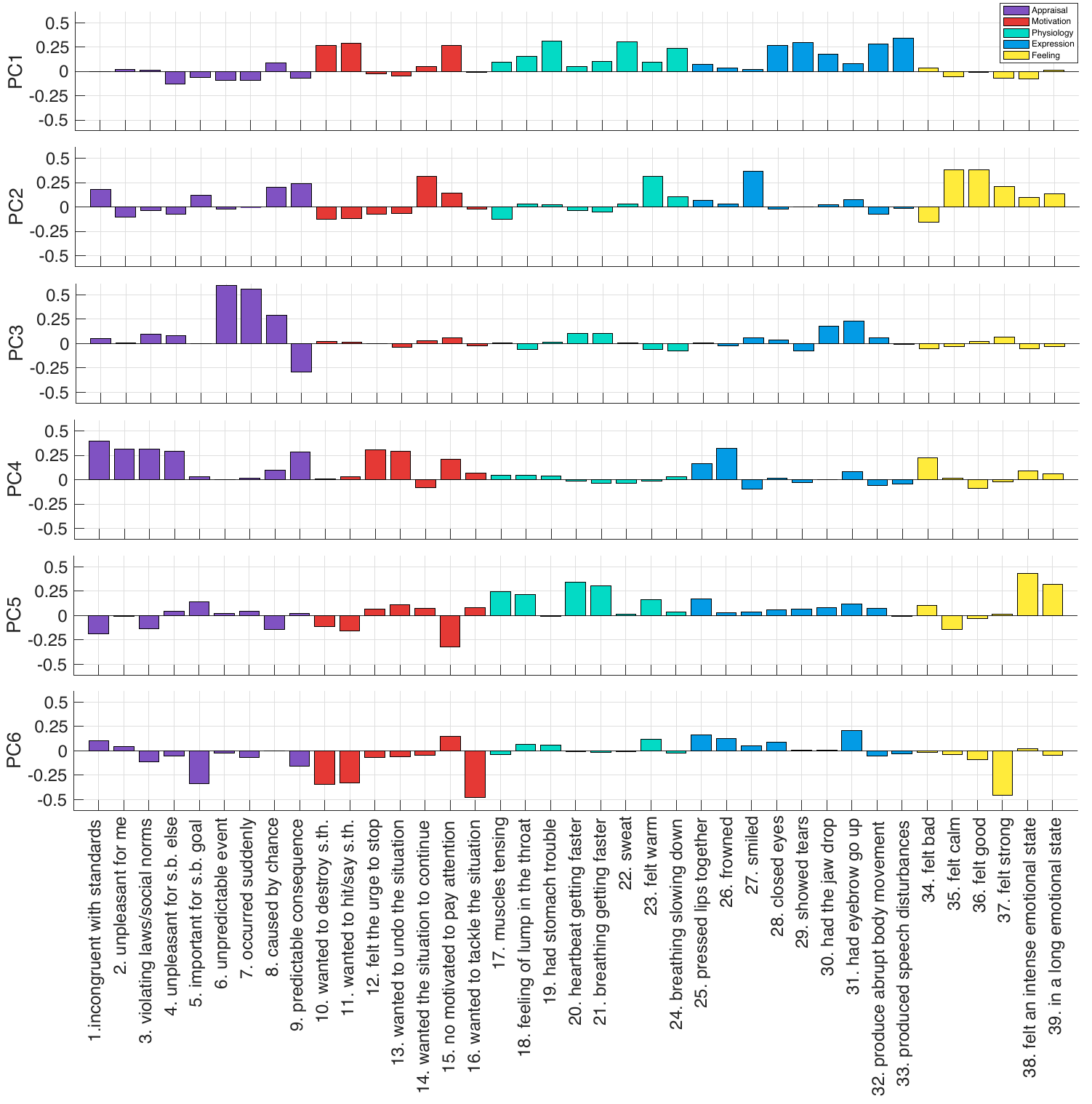}
\end{center}
\caption{Coefficients of the first six principal components of GRID features. The features of different components are colour-coded differently. To interpret each sub-dimension (principal component ), items are considered according to their weights. We interpret these dimensions as action tendency, pleasant feeling, novelty, valuation of norms, arousal and goal relevance, respectively. }\label{componentAnalysis}
\end{figure*}

\subsection{Dimensional Analysis}

In the second step to reveal the main factors underpinning an emotional experience with greatest variance, we applied a Principal Component Analysis (PCA) to the GRID features\cite{Wold1987}. We selected the first six principal components which accounted for about 59\% of the total variance. The reason for choosing only six components was due to gaining little variance by retaining more components. A Varimax rotation was applied to simplify the interpretation of each sub-dimension in terms of just few major items. The interpretation of these six dimensions is based on their relationships with the GRID features. Figure~\ref{componentAnalysis} demonstrates the coefficients for each of the six components. The first component ($30\%$ of total variance) loads mostly on items related to motivation (e.g. tendency to destroy something or say something), expressions (e.g closing eyes, showing tears) and changes in physiology which can be interpreted as action tendency. The second component ($14\%$ of total variance) which is mainly correlated with items in feeling component (e.d feeling good, not feeling bad, feeling calm), smile and feeling warm seems to encode pleasant feeling (vs. unpleasant feeling). The third component ($5\%$ of total variance) encodes appraisal of suddenness and unpredictability that together can be interpreted as novelty checks. The forth component ($4\%$ of total variance) is heavily loaded on appraisal items related to violation of norms and standards which is unpleasant for self and others, that can represent the valuation of norms. The fifth component ($4\%$ of total variance) has high correlation with long and intense emotional experience with high breathing rate and fast heartbeat that together represent the arousal state. The sixth component ($3\%$ of total variance) which has been characterised by a relatively high loadings on relevance for somebody's goal in appraisal component and high motivation for taking some actions without high loadings in body and physiological changes that comes with feeling strong can be interpreted as appraisal of goal relevance. These factors are in line with the findings of previous studies based on similar component models~\cite{Fontaine2015}, and confirm that in order to characterise different emotional experience, more than two dimensions of valence and arousal are needed to capture the commonality and specificity of different types of emotions. 

\subsection{Modelling}

The last step consists in modelling and predicting the categorical emotion label from the GRID features. To evaluate the capacity of using GRID features to predict the categorical emotions, first we simplified the problem to a binary classification. To do so, we grouped the ratings for each categorical emotion into two classes of \emph{``high"} (equal or above the median) and \emph{``low"} (below the median). Initially, one \emph{Logistic Regression} (LR) classifier per each categorical emotion was trained and tested using k-fold cross validation method with $k = 10$. At each iteration of training, one fold was left out as the test set to evaluate the generalisability of the model to unseen data. The folds were selected such that the samples from one assessor appear only in one fold to ensure that the training and test sets are independent. To evaluate nonlinear relationships and the effect of feature interactions, further models were also applied including \emph{Random Forest} and nonlinear \emph{Support Vector Machine} (SVM) with gaussian kernel. Although, results showed marginal improvement comparing to \emph{Logistic Regression} and linear SVM, the difference is not statistically significant and it is hard to conclude that nonlinear models can capture the relationship between GRID features and discrete emotions, better.
\begin{figure}[!t]
\begin{center}
\includegraphics[width=9cm]{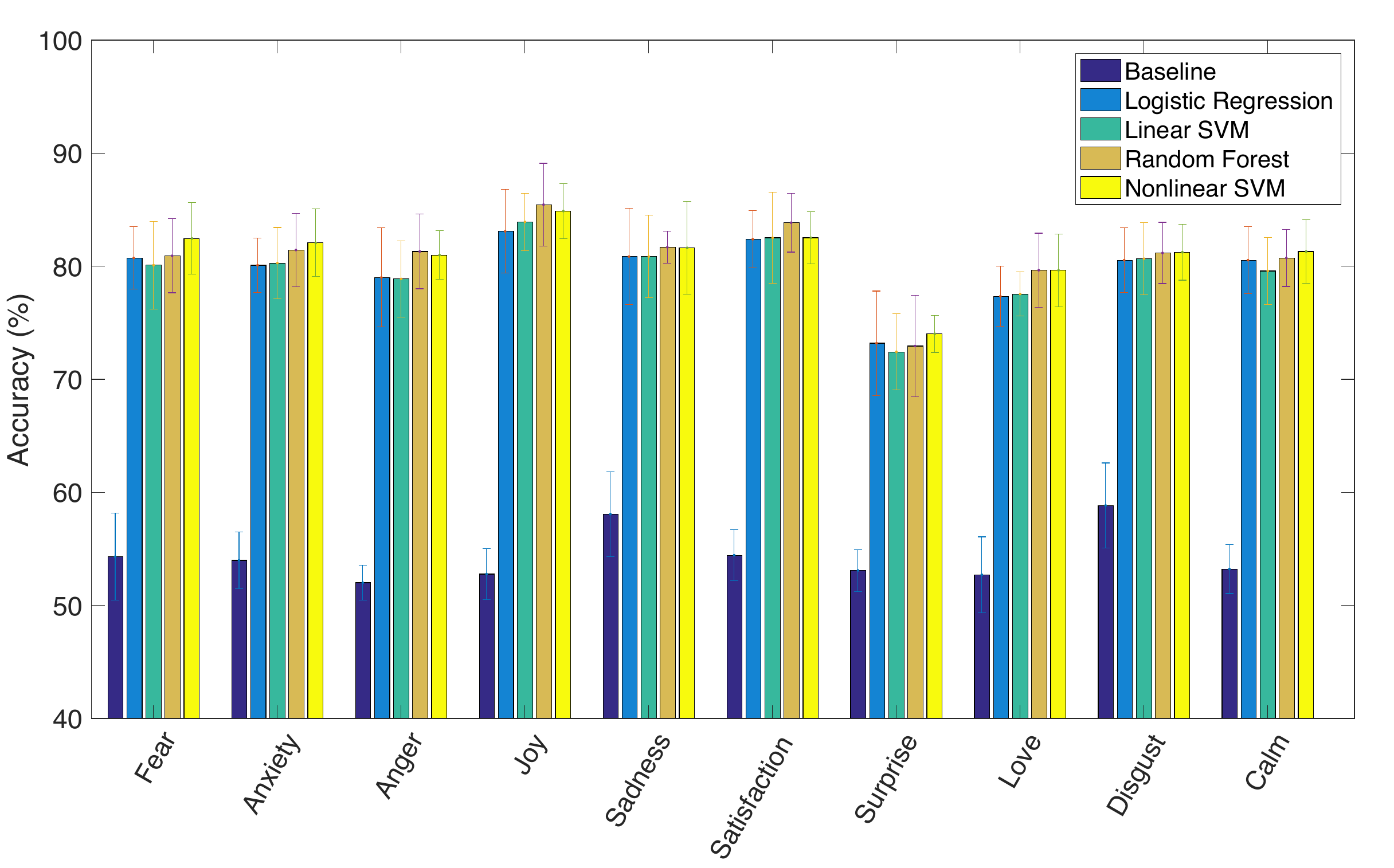}
\end{center}
\caption{Accuracy of different models for binary classification of the discrete emotions using GRID features. Error bars represent standard deviation of the accuracy from 10-fold cross validation.}\label{BinaryClassification}
\end{figure}

Figure~\ref{BinaryClassification} shows the accuracy of the binary classifications for each category and the corresponding baseline. For all categories and all different models, accuracy is significantly higher than baseline ($p<0.001$), however the best performances are for joy, satisfaction, calm and sadness in which the first three share very similar characteristics given that the calm category in our dataset comes with more positive valence (see Figure~\ref{ClusterAnalysis}).

\begin{table*}[!htbp]
\begin{center}
\begin{tabular}{|p{4cm}|p{0.8cm}|p{0.8cm}|p{0.8cm}|p{0.8cm}|p{0.8cm}|p{0.8cm}|p{0.8cm}|p{0.8cm}|p{0.8cm}|p{0.8cm}|}
	\hline
	                   &  Fear  & Anxiety & Anger & Joy & Sad & Satisfied & Surprise & Love & Disgust & Calm\\                                                                                                                                 
	\hline
		\textbf{Appraisal}& $1.08$  & $1.01^*$ & $1.21$ & $1.54$ & $1.31$ & $1.29$ & $1.04^{**}$ & $1.49$ & $0.98^*$  & $1.2$ \\
	\hline
		\textbf{Motivation} & $1.07^*$ & $0.97^{**}$ & $1.07^*$ & $1.45$ & $1.18$ & $1.16$ & $1.26$ & $1.52$ & $1.09^*$ & $1.09$\\
	\hline
		\textbf{Physiology} & $0.92^{**}$ & $0.88^{**}$ & $1.33$ & $1.43$ & $1.30$ & $1.29$ & $1.29$ & $1.23$ & $1.29$ & $1.13$ \\
	\hline	
		\textbf{Expression} & $1.04^{**}$ & $0.91^*$ & $1.09^*$ & $0.81^{**}$ & $1.15$ & $0.86^{**}$ & $1.13$ & $1.04^*$ & $1.09$ & $1.14$\\
	\hline
		\textbf{Feeling} & $0.95^{**}$ & $0.86^{**}$ & $1.06$ & $0.98^*$ & $0.92^{**}$ & $0.82^{**}$ & $1.22$ & $0.99^*$ & $1.10$ & $0.80^{**}$ \\
	\hline
	\hline
		\textbf{All Components except Feeling} & $0.80^{**}$ & $0.74^{**}$ & $0.91^{**}$ & $0.76^{**}$ & $0.95^{**}$ & $0.79^{**}$ & $0.92^{**}$ & $0.92^{**}$ & $\textbf{0.89}^{**}$ & $0.94^{**}$\\

	\hline
		\textbf{All Components} & $\textbf{0.77}^{**}$	 & $\textbf{0.73}^{**}$ & $\textbf{0.83}^{**}$ & $\textbf{0.70}^{**}$ & $\textbf{0.83}^{**}$ & $\textbf{0.69}^{**}$ & $\textbf{0.89}^{**}$ & $\textbf{0.86}^{**}$ & $\textbf{0.89}^{**}$ & $\textbf{0.78}^{**}$\\

	\hline
	\end{tabular}
\end{center}
\caption{Performance of \emph{Ordinal Regression} for prediction of each discrete emotion ranking from GRID features of one component or all components together. Numbers represent $MAE^M$ and the baseline $MAE^M$ for trivial ordinal ranking model is $1.2$ for all OR models. Asterisks show level of significance where $^*$ means $p<0.01$ and $^{**}$ means $p<0.001$. Numbers in bold show the best performance for each discrete emotion.}\label{classificationResult}
\end{table*}

The binary classification results demonstrated the capacity of GRID features in making distinction between \emph{high} and \emph{low} values of each categorical emotion. However, the assessments of discrete emotions are in the form of qualitative ratings of individual items (e.g., from 1 for "not at all" to 5 for "strongly"). Although no explicit measure of distance can be defined between adjacent categories, the ratings possess properties of ordinality (e.g., "to some extent" $>$ "not at all"). Therefore to find the mapping between each discrete emotion rating and its representation in GRID space, the more appropriate approach is to apply some form of preference learning. In this work, we used \emph{Ordinal Regression} based on \emph{Proportional Odds} model.

\emph{Ordinal Regression} (OR) is a well suited approach for automatically predicting an ordinal variable\cite{Gutierrez2016}. 
In OR, a set of $n$ samples $\{\xvect_{1},\ldots,\xvect_{n}\}$, each described by $d$ features ($\xvect_i \in \mathbb{R}^d$), is associated with a set of labels $\{y_{1},\ldots,y_{n}\}$ selecting one of $r$ ordered categories in $C = \{1,\ldots,r\}$ representing the ranking of the corresponding inputs.
Let $\yvect = (y_1, \ldots, y_n)^{\T}$ and $X$ be the $n \times d$ matrix obtained by stacking the input vectors by row.

\emph{Proportional Odds Model} (POM) uses cumulative probabilities as follows:
\begin{equation}
  \log\left[\frac{p( y \leq h | \xvect)}{p(y > h | \xvect)} \right] = \alpha_h + \xvect^{T} \betavect,
 \end{equation}
which assumes that the logarithm of proportional odds on the left hand side can be expressed as a linear combination of covariates with parameters $\betavect$ and a bias term
$\alpha_h$ which depends on $h$, with $\alphavect = (\alpha_1, \ldots, \alpha_{r})$.
It can be shown that the above is equivalent to:
\begin{equation}
 p(y \leq h | \xvect,) = \frac{1}{1+\exp[-(\alpha_{h}+\xvect^{\T}\betavect)]}=l(\alpha_{h}+\xvect^{T}\betavect)
\end{equation}
where $l(z)$ is the \emph{logistic function} and the probabilities for the observed $y_i$ can be obtained from $p(y_i=h|x_i) = l(\alpha_{h} + \xvect_i^{T} \betavect) - l(\alpha_{h-1}+ \xvect_i^{T} \betavect)$ for $h>1$ and $p(y_i=h)=l(\alpha_{1} + \xvect_i^{T} \betavect)$ for $h=1$. The parameters of the model $\betavect$ are estimated using \emph{Maximum Likelihood} and can be used to interpret the relation between features and the response variable.

For each discrete emotion category we first performed five OR models to predict the rating from one component at the time (e.g. appraisal, expression, etc.) to evaluate the predictive power of different components for each discrete emotion, separately.  Then we predicted emotions once using all five components (all GRID features) and next from all components except \emph{feeling }to assess the strength of components that do not directly measure the feeling in prediction of discrete emotions, which is a more plausible scenario in automatic emotion recognition.
A 10-fold cross validation procedure (similar to above) was performed and to evaluate the performance of the model we used macro-averaged mean absolute error (denoted as ${MAE}^M$) which is a modified version of mean absolute error  ($MAE$) to account for imbalanced data classes. $MAE$ is the average of absolute deviation of predicted rank $y_i^*$ from the actual rank $y_i$: and ${MAE}^M$ is the average of absolute error across classes:
\begin{equation}
	MAE^M = \dfrac{1}{n} \sum_{j=1}^{r} \dfrac{1}{T_j} \sum_{x_i\in T_j}{}|y_{i}-y_{i}^*|
\end{equation}
\begin{figure*}[!htbp]
\begin{center}
\includegraphics[width=18.5cm]{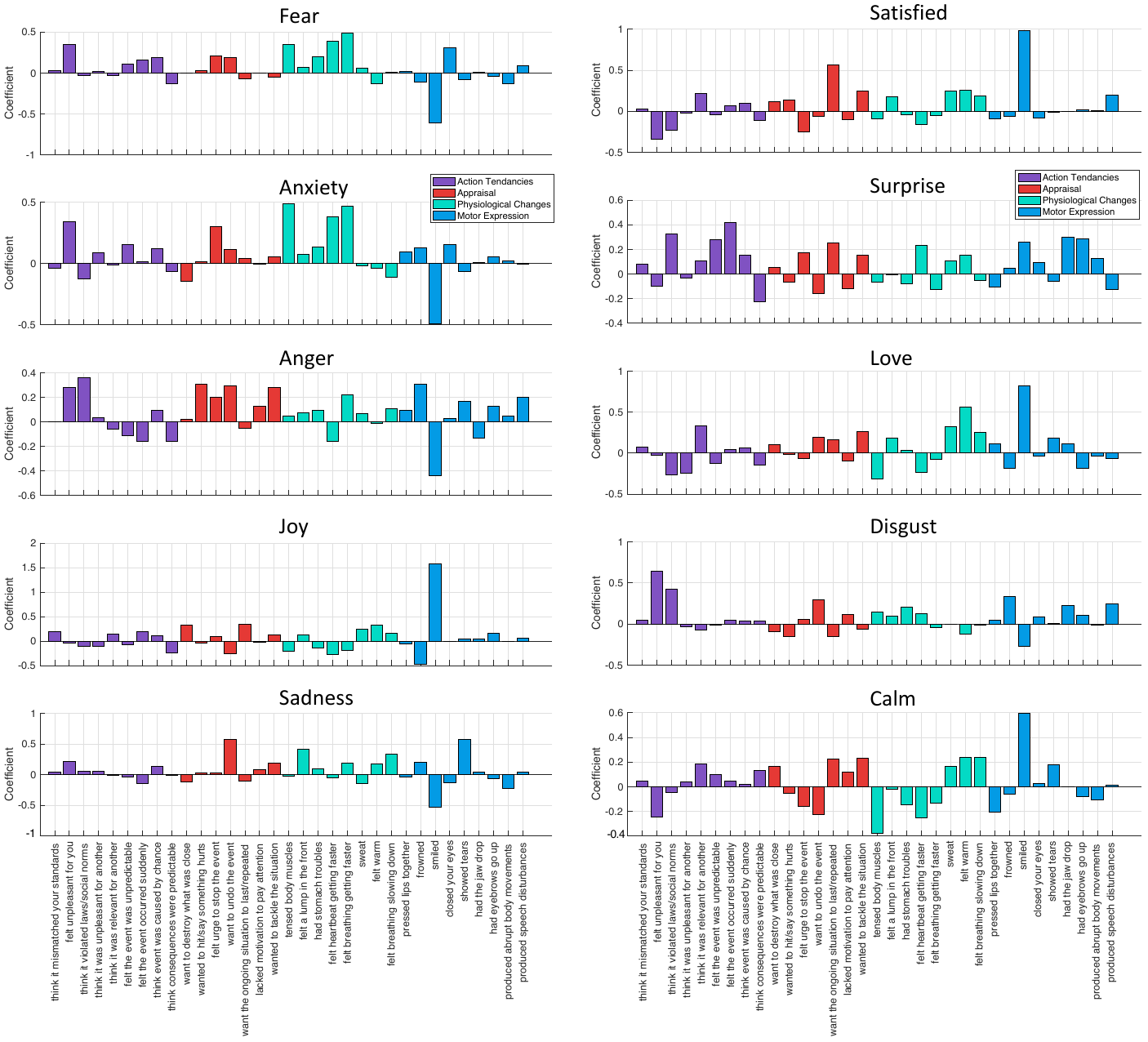}
\end{center}
\caption{Coefficients of the OR model for each discrete emotion. Features from different components are colour-coded differently. Please note that, for the purpose of better visualisation,  the scale of y-axis is different for different emotions.}\label{OR_Coeff}
\end{figure*}

where $T_j$ is the set of samples, $x_i$, whose true rank is $j$. Although $MAE$ is the most common measure for evaluating ordinal regression models, it is not robust against imbalanced dataset. Therefore, we use ${MAE}^M$ which is a more rigorous metric to account for imbalanced ordinal classes\cite{Baccianella2009}. Table~\ref{classificationResult} reports the evaluation of the model on each discrete emotion. The results suggests that although \emph{expression}, the most dominant component in the field of \emph{automatic emotion recognition}, is a powerful component to predict most of the categories, it may not be as discriminative for some of the categories such as sadness, surprise, disgust and calm (at least in this context). This means that such categories can be more of an internal state without any external expression. However, for sadness and calm, \emph{feeling} component which mainly captures valence and arousal in our experiment is more predictive. For surprise and disgust, \emph{appraisal} component is the most informative component due to capturing the internal evaluation of unpredictability and anti-sociality. \emph{Physiology} component shows high distinction power for fear and anxiety and low distinction power for joy. As expected \emph{motivation} component is more discriminative for fear, anxiety, anger and disgust, where people usually show more action tendencies to tackle or stop the situation. Using all components result significant improvements in all emotion categories emphasising the importance of taking the full componential model into account. And finally, removing \emph{feeling} component leads to a decrease in ${MAE}^M$ in almost all emotions, except disgust, when compared against using all components. The largest gap appears in calm, followed by sad and satisfied which indicate the importance of \emph{feeling} component in prediction of those emotions. All together, these findings suggest that no single component is enough to capture the difference of all categories and adding components like appraisal and motivation which can be captured to some extent by including context can potentially improve the results significantly. This finding is similar to what was reported in \cite{Meuleman2018} that a subset of features from all components resulted the best performance for joy and fear, however at individual component level, our finding is in contrast to what they found in which \emph{appraisal} was the most predictive component for joy and fear. This difference can be partially explained by the different emotion elicitation procedures, active vs passive where the former has higher personal relevance and subsequently stronger appraised concerns.

\begin{table*}[ht]
\begin{center}
\begin{tabular}{|p{2cm}|p{1cm}|p{1cm}|p{1cm}|p{1cm}|p{1cm}|p{1cm}|p{1cm}|p{1cm}|p{1cm}|p{1cm}|}
	\hline
	                   & \begin{sideways} \textbf{Felt bad}  \end{sideways} & \begin{sideways} \textbf{Felt calm } \end{sideways} & \begin{sideways} \textbf{Felt good} \end{sideways}& \begin{sideways} \textbf{Felt strong} \end{sideways} & \begin{sideways} \textbf{Intense emotion} \end{sideways} & \begin{sideways} \textbf{Lasting state} \end{sideways}  \\                                                                                                                                 
	\hline
		\textbf{Extraversion}& $ -0.02$  & $+0.04$ & $+0.10^{**}$ & $ +0.16^{**}$ & $+0.00 $ & $+0.05$  \\
	\hline
		\textbf{Agreeableness}& $ +0.01$ & $-0.03$ & $+0.02$ & $ -0.03$ & $-0.03 $ & $+0.02$  \\
	\hline
		\textbf{Conscientiousness}& $ -0.04$  & $-0.02$ & $-0.04$ & $ +0.04$ & $+0.00 $ & $-0.04$  \\
	\hline
		\textbf{Neuroticism}& $ +0.08^{*}$  & $-0.08^{*}$ & $-0.06$ & $ -0.14^{**}$ & $+0.05 $ & $+0.05$  \\
	\hline
		\textbf{Openness}& $+0.05$  & $-0.07^{*}$ & $-0.18^{**}$ & $ -0.14^{**}$ & $-0.03 $ & $-0.12^{**}$  \\
	\hline

	\end{tabular}
\end{center}
\caption{Partial correlation between personality and feeling component. Asterisks indicate level of significance where $^*$ corresponds to $p<0.01$ and $^{**}$ corresponds to $p<0.001$}\label{PersonalityFeelingCorrelation}
\end{table*}
\begin{table*}[ht]
\begin{center}
\begin{tabular}{|p{2cm}|p{1cm}|p{1cm}|p{1cm}|p{1cm}|p{1cm}|p{1cm}|p{1cm}|p{1cm}|p{1cm}|p{1cm}|}
	\hline
	                   &  \textbf{Fear}  & \textbf{Anxiety} & \textbf{Anger} & \textbf{Joy} & \textbf{Sad} & \textbf{Satisfied} & \textbf{Surprise} & \textbf{Love} & \textbf{Disgust} & \textbf{Calm}\\                                                                                                                                 
	\hline
		\textbf{Extraversion}& $ -0.06$  & $-0.08^{*}$ & $-0.01$ & $ +0.06$ & $+0.01 $ & $+0.06$ & $+0.07$ & $+0.07^{*}$ & $-0.07^{*}$  & $+0.07^{*}$ \\
	\hline
		\textbf{Agreeableness} & $+0.01$ & $+0.02$ & $+0.06$ & $+0.01$ & $+0.02$ & $-0.01$ & $+0.01$ & $+0.02$ & $+0.06$ & $-0.06$\\
	\hline
		\textbf{Conscientiousness} & $-0.09^{**}$ & $-0.07^{*}$ & $-0.12^{**}$ & $-0.09^{**}$ & $-0.11^{**}$ & $-0.10^{**}$ & $-0.01$ & $-0.09^{**}$ & $-0.10^{**}$ & $+0.00$ \\
	\hline	
		\textbf{Neuroticism} & $+0.12^{**}$ & $+0.16^{**}$ & $+0.13^{**}$ & $-0.04$ & $+0.09$ & $-0.02$ & $-0.03$ & $-0.01$ & $+0.11^{**}$ & $-0.12^{**}$\\
	\hline
		\textbf{Openness} & $-0.02$ & $-0.05$ & $-0.11^{**}$ & $-0.13^{**}$ & $-0.10^{**} $ & $-0.15^{**}$ & $-0.09^{**}$ & $-0.13^{**}$ & $-0.02$ & $-0.10^{**}$ \\
	\hline

	\end{tabular}
\end{center}
\caption{Partial correlation between personality and discrete emotions. Asterisks indicate level of significance where $^*$ corresponds to $p<0.01$ and $^{**}$ corresponds to $p<0.001$}\label{PersonalityAffectCorrelation}
\end{table*}

To evaluate the importance of each GRID features in the prediction of different emotions, we analysed the $\betavect$ coefficients of the OR model. Coefficients with higher values, have higher impact in the classification of a particular emotion from the GRID features (see Fig.~\ref{OR_Coeff}) . Analysing coefficients shows that one of the most important features for almost all discrete emotions is smile, which shows positive relation with positive emotions and negative relation with negative emotions as expected. This suggest that smile is a good indicator of the emotion valence at a high level. Features related to appraisal seem more informative for surprise, disgust, anxiety and anger. Appraisals of unpleasantness, violation of social norms and suddenness indicate higher relevance for different emotions. As expected, a number of features in \emph{motivation} component are relevant for anger which are indicating tendency to take some actions. This is opposite to calm that shows lack of tendency to take any action. For sadness, undoing the event is the most important features while for surprise and satisfied continuing the ongoing situation is among the more relevant. The \emph{physiology} component is more relevant for anxiety and fear with tensed body muscle, heart rate and respiratory rate among the most important features to rank them. Feeling warm is an informative feature for love, and relaxed body muscle (opposite of tensed body muscle) is a good indicator for calmness. In \emph{expression} component, beside smile, frowning is a good indicator for anger and disgust, showing tears is an informative sign of sadness, eyebrows going up and jaw drop are mostly relevant in surprise and finally producing speech disturbances is more an indicator of disgust.

\begin{figure*}[!htbp]
\begin{center}
\includegraphics[width=16cm]{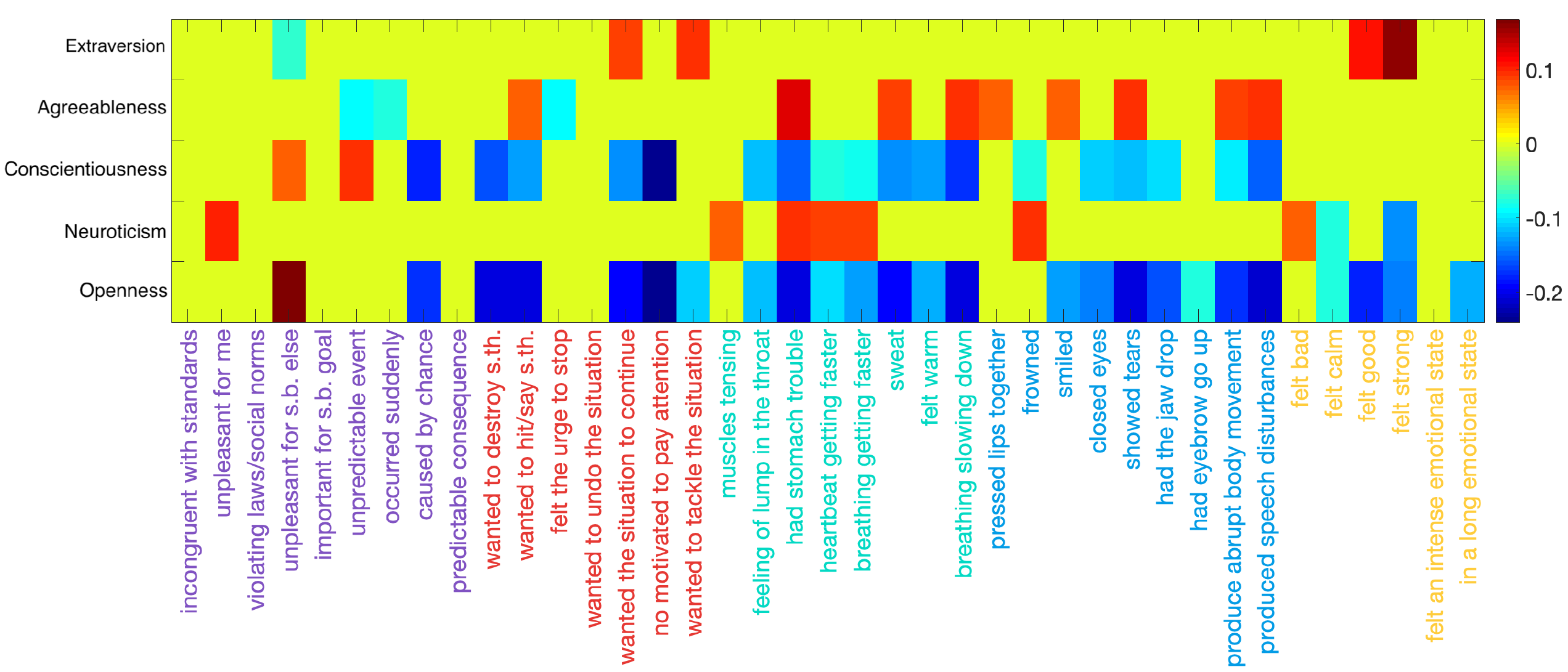}
\end{center}
\caption{Partial correlation between personality and all CoreGRID items; For the purpose of better visualisation, only statistically significant coefficients with $p<0.01$ are highlighted and the rest are set to zero. CoreGRID items are colour coded according to the component they belong to.}\label{Personality_CoreGRID}
\end{figure*}

\subsection{Personality-Emotion Relationship}
To examine the relationship between personality and emotion and how personality may affect the level of experienced emotion, we conducted three correlation analyses: the first one between personality and questions related to feeling component (that mainly encode for valence and arousal) so that would be directly comparable with the results in majority of literature that have mainly focused on dimensional model, the second one between personality and discrete emotions and the last one between personality and component model descriptors. We postulate that feelings and discrete emotion correlations reflect the individual differences in emotional experience resulting from variations in appraisals of external events. As participants have reported the emotion across different stimuli, and the assessments are in the form of rankings (likert-scale), we applied partial Spearman correlation. Table~\ref{PersonalityFeelingCorrelation} and Table~\ref{PersonalityAffectCorrelation} report the partial correlation coefficients between personality traits and, feeling component and discrete emotions, respectively. No correction has been applied to the correlation coefficients. 

The feeling component has been described through six items including: feeling bad, feeling calm, feeling good, feeling strong, feeling an intense emotion and experiencing a lasting state. The correlation analysis indicates significant correlations between \emph{Neuroticism} and feeling bad and feeling calm indicating that people with higher levels of neuroticism feel higher levels of negative valence and less calm. However, such correlation was not found for feeling good which potentially suggests that \emph{Neuroticism} does not have significant correlation with valence in positive stimuli. This is partially in contrast to our initial hypothesis and is more in line with the findings of  \cite{Ng2009}, \cite{Stenberg1992} and \cite{Pascalis2004} who also could not find a strong correlation in positive stimuli. \emph{Extraversion} showed positive correlations with feeling good and feeling strong which imply a relationship with positive affect and confirms our initial hypothesis, but no correlation was found for feeling an intense emotion or experiencing a lasting state that together can encode for arousal. \emph{Conscientiousness} and \emph{Agreeableness} did not show any correlation with the items from feeling component, which is not in line with our initial hypotheses. And finally, \emph{Openness}  indicates a significant correlation with multiple items and shows a negative significant correlation with feeling good which is also opposite to our hypothesis and the findings of a meta-analysis study in~\cite{Steel2008}. In addition \emph{Openness} showed a significant negative correlation with experiencing intense state, suggesting that those higher on this trait have experienced shorter emotional states. No significant correlation was found between feeling an intense emotion and any of personality traits in this study.

In the second analysis, we looked at the relationship between personality and discrete emotions to examine the effect of personality traits on experiencing a specific emotion. We hypothesised that the individual differences in the experienced discrete emotions cannot all be captured by the items related to the feeling component (which only captures valence and arousal changes), and the effect might be different across different discrete emotions even with similar dimensional model representations (e.g fear and anger).  For \emph{Extraversion} we found only marginal relationships with anxiety, love, disgust and calm and the direction of association is consistent with what we observed in the first analysis. No significant correlation was found for \emph{Agreeableness} which is in accordance with the first analysis as well. Similar to the first analysis \emph{Neuroticism} shows significant positive correlations with all negative emotions and significant negative correlation with calm without any significant correlation with pleasant emotions, implying that highly neurotic people experience higher levels of unpleasant emotions. 

Overall, \emph{Openness} and \emph{Conscientiousness} showed a larger number of significant correlations with discrete emotions. In line with the first analysis all emotions with positive valence (e.g joy, satisfied and love) have significant negative correlation with \emph{Openness}; however, negative valence emotions such as anger and sadness show a significant negative correlation with this trait; Although this might seem incompatible with our first correlation analysis which did not show any significant correlation with feeling bad, we believe this is due to the effect of personality on other dimensions of emotion such as action tendency or novelty that are not captured by the feeling component. Similarly, \emph{Conscientiousness} that did not have any significant correlation with any of the feeling items in the first analysis, shows significant negative correlations with all discrete emotions except surprise and calm, suggesting that people with high score for these traits may experience lower levels of emotions in general or it might be more difficult to elicit emotions in them. Again, we hypothesise that this seemingly inconsistent result is due to the effect of personality on other aspects of emotion like novelty and action tendencies that are not captured in the feeling component. In order to examine our hypothesis regarding the effect of personality on other dimensions of emotion, we also looked at the correlation between personality trait scores and the CoreGRID descriptors used in this study. Figure~\ref{Personality_CoreGRID} depicts the correlation coefficients and for the purpose of better visualisation, only those values that were statistically significant with $p<0.01$ were highlighted. The result suggests that as hypothesised, personality can affect aspects of emotion other than feeling pleasant/unpleasant and arousal. It is particularly evident that the two personality traits, \emph{Conscientiousness} and \emph{Openness} affect the items related to motivation, physiology and expression more than the other traits which can help to justify the incompatibility between the result from first analysis and the second one. In addition the strength of correlation between personality traits and some of the CoreGRID descriptors are much stronger than those found between personality and discrete or dimensional emotions/

In addition, in a separate analysis, similar to the analysis in \emph{Modelling} section , we included the personality scores along with all other components to predict discrete emotions, but no improvement in the result was obtained. We believe that this is due to the fact that emotion components can capture the individual differences reflected in the personality scores as shown in the correlation analysis above.

\section{Conclusion}
This paper analysed the relationship between discrete emotions and componential model of emotion which postulates five major components underlying emotional experience: appraisal, motivation, physiology, expression and feeling. A set of 1576 surveys of emotional experience assessment was used in the analysis. The survey included a subset of 39 GRID features that evaluates the changes in each of the five components along with an assessment of 10 discrete emotions that participants had to respond after watching video clips with emotional content. To the best of our knowledge this is among the first studies that evaluated the relationship between discrete emotions and component model of emotion using all five components. Moreover, previous studies have mostly focused on empirical assessment of semantic profile of emotion whereas this study has focused on the assessment of actual emotional experience using a data-driven approach.

Four types of analyses were carried out starting by clustering the discrete emotion profiles in componential space, which yielded a clear distinction between positive and negative emotions in the high level and separate clusters of happiness, serenity, distress, anger and surprise in the low level. In the second analysis, we performed a dimensional reduction technique to reveal the most important dimensions underlying the emotional experience. Six dimensions were retained that represent action tendency, pleasantness, novelty, valuation of norms, arousal and goal relevance. These dimensions suggest that more than two dimensions of valence and arousal are needed to distinguish between different types of emotional experience. Next, we predicted the rating of each discrete emotion category from GRID features, first by utilising one component at a time and then combining all components. The results are higher than chance level with high statistical significance for all discrete emotions if we use all components, however using individual components decrease the performance significantly in all categories. The results also suggest that different components contribute differently to the prediction of each emotion category. These findings highlights the importance of taking a multi-componential approach in recognition of emotions. Finally, we analysed the personality-affect relationship and observed negative correlation between \emph{Neuroticism} and valence in negative stimuli. In addition, \emph{Openness} and \emph{Conscientiousness} showed mostly negative correlations with different emotion categories, suggesting lower level of emotions for people who have high scores in those traits. Moreover, we showed that personality traits are correlated with some of the descriptors of different components, suggesting that personality dispositions can potentially affect our perception and reaction to an emotional experience by changing appraisal and motivational components followed by the physiological and expressional changes. 
 
There are also some limitations in the current study to be acknowledged. First, the emotional assessments for each clip is done at global level which neglects the presence of different events in the clip with potentially different emotional content. Therefore, one direction for future work is to limit the emotional assessment to shorter episodes to have a more fine-grained assessment. The second limitation is the passive nature of emotion elicitation in this study, in which participants are not actively involved in an event with potential implication for oneself. Therefore, it is important to note that the findings of this study, at best, applies to passive emotional experience. Virtual reality or interaction scenarios which are more immersive might be suitable to overcome such a limitation. The third limitation is collecting data through online platforms rather than a controlled environment which potentially introduces more noise to the data. And finally, using self-reports for components such as \emph{physiology} and \emph{expression} that people do not continuously and consciously monitor may not be very accurate and objective recording are more desirable.

\section*{Acknowledgment}
This work was supported by the Swiss National Science Foundation through Sinergia grant number 180319 and National Centre of Competence in Research (NCCR) Affective Sciences. Authors would like to thank Ma\"elan Men\'etrey for his help with data collection and  Kangying Lin for preliminary analysis.

\bibliographystyle{IEEEtran}
\bibliography{library}  			

\begin{IEEEbiography}[{\includegraphics[width=1in,height=1.25in,clip,keepaspectratio]{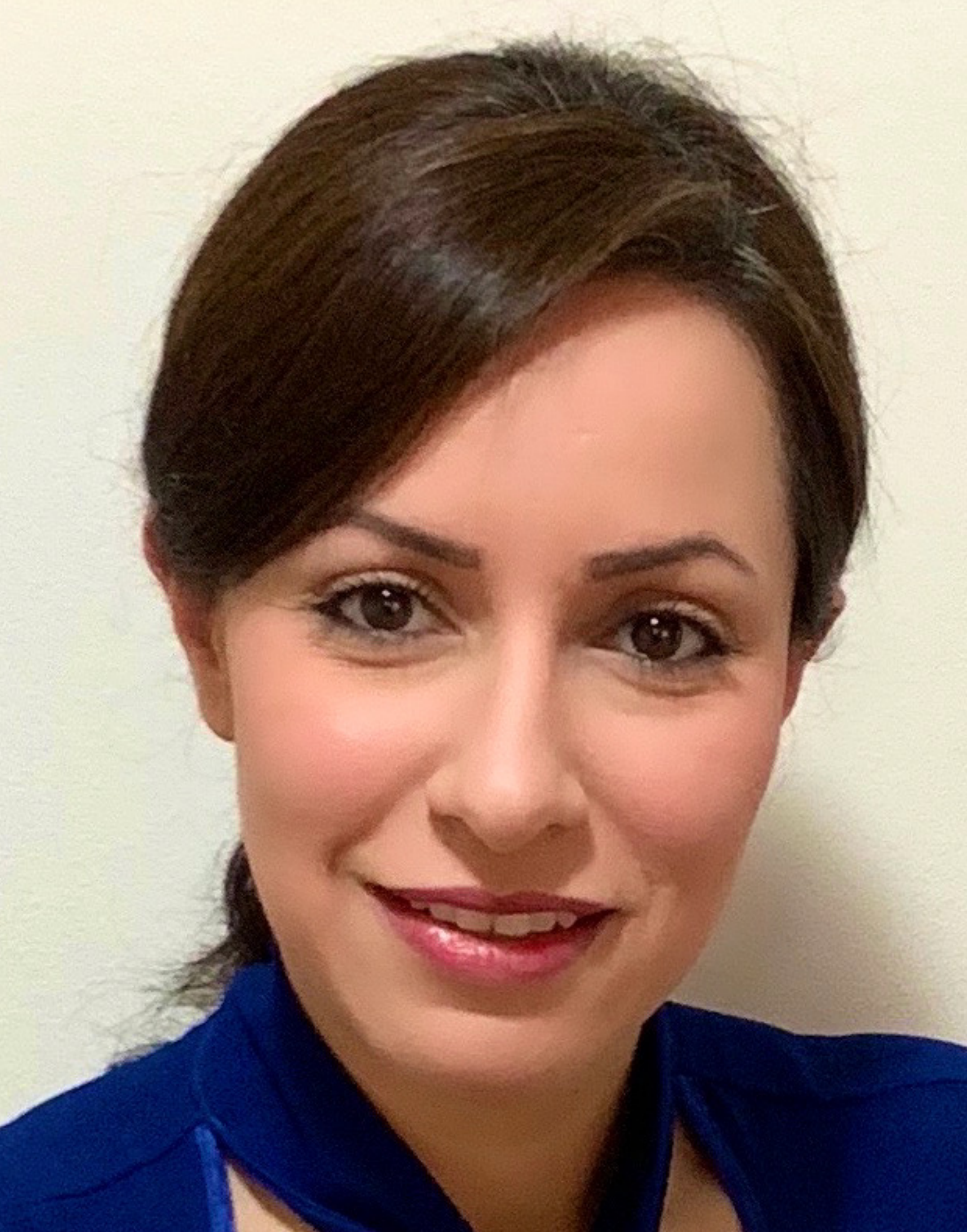}}]{Gelareh Mohammadi} 
(Member, IEEE) received the PhD in electrical engineering from EPFL, Switzerland, in 2013. She is with the University of New South Wales, Australia, where she is a lecturer (assistant professor) at the School of Computer Science and Engineering. Her main research interests include human behavior Analysis and Social Signal Processing, and their applications in psychology and medicine. She was a post-doctoral researcher at Idiap Research Institute until 2014 investigating how people perceive others personality from their voice. Then she joined the Laboratory for Neurology and Imaging of Cognition, University of Geneva,  as a postdoctoral researcher where she was working on understanding neural basis of emotion using a data-driven approach and remained there until 2018 before she joined UNSW. She has  authored and co-authored more than 40 publications, including several book chapters. She was a recipient of the "Google Anita Borg award" in 2013 and has been selected as the Australia's research field leader in human-computer-interaction in the Australian Research Report, 2019.

\end{IEEEbiography}

\begin{IEEEbiography}[{\includegraphics[width=1in,height=1.25in,clip,keepaspectratio]{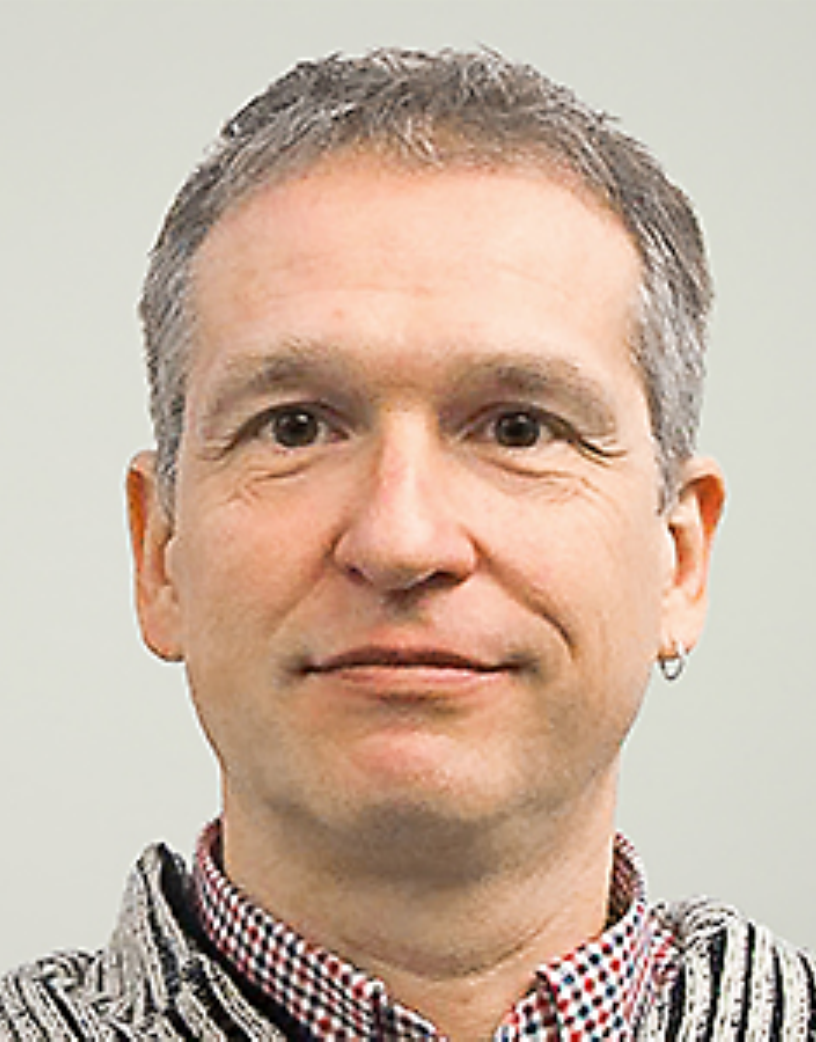}}]{Patrik Vuilleumier} 
is a neurologist who is using brain imaging techniques (such as functional resonance magnetic imaging, fMRI) to study the cerebral mechanisms of emotion and awareness. After first training in clinical neurology and neuropsychology in Geneva, Lausanne, and Paris, he pursued his research in cognitive neurosciences at the University of California in Davis, and then at University College London. He is now heading the Laboratory for Neurology and Imaging of Cognition at Geneva University Medical Center and Hospital. He is also director of the Interdisciplinary Neuroscience Center at the University of Geneva since 2007 and co-director of the neuroimaging center created by the University of in Geneva since 2008 (Brain \& Behaviour Laboratory). He is both a research leader and steering board member of the Swiss Center for Affective Sciences, a multidisciplinary national institute devoted to research on emotions in individuals and societies, supported by the Swiss government and hosted by the University of Geneva. For more than 15 years, Patrik Vuilleumier has studied how emotion processing can influence perception, attention, and action, using functional neuroimaging techniques (fMRI) and neuropsychological studies of brain-damaged patients. His current work concerns the development of imaging research and applications for investigations in neurological and psychiatric diseases in Geneva Medical School, in addition to more basic research on emotion processes in the normal brain. He authored more than 200 publications and obtained several awards for his work, including the "Bing Prize in Neuroscience" (2004) from the Swiss Academy of Medical Sciences, and the "Distinguished Scientific Award for Early Career Contribution in Behavioral and Cognitive Neuroscience" (2007) from the American Psychological Association (APA).
\end{IEEEbiography}

\end{document}